\documentclass[useAMS,usenatbib,epsfig]{mn2e}

\newif\ifAMStwofonts
\AMStwofontstrue
\usepackage{amsmath,amsfonts,epsfig,natbib}

\def\reff@jnl#1{{\rm#1\/}}

\def\aj{\reff@jnl{AJ}}                  
\def\araa{\reff@jnl{ARA\&A}}            
\def\apj{\reff@jnl{ApJ}}                        
\def\apjl{\reff@jnl{ApJ}}               
\def\apjs{\reff@jnl{ApJS}}              
\def\ao{\reff@jnl{Appl.Optics}}         
\def\apss{\reff@jnl{Ap\&SS}}            
\def\aap{\reff@jnl{A\&A}}               
\def\aapr{\reff@jnl{A\&A~Rev.}}         
\def\aaps{\reff@jnl{A\&AS}}             
\def\azh{\reff@jnl{AZh}}                        
\def\baas{\reff@jnl{BAAS}}              
\def\jrasc{\reff@jnl{JRASC}}            
\def\memras{\reff@jnl{MmRAS}}           
\def\mnras{\reff@jnl{MNRAS}}            
\def\pra{\reff@jnl{Phys.Rev.A}}         
\def\prb{\reff@jnl{Phys.Rev.B}}         
\def\prc{\reff@jnl{Phys.Rev.C}}         
\def\prd{\reff@jnl{Phys.Rev.D}}         
\def\prl{\reff@jnl{Phys.Rev.Lett}}      
\def\pasp{\reff@jnl{PASP}}              
\def\pasj{\reff@jnl{PASJ}}              
\def\qjras{\reff@jnl{QJRAS}}            
\def\skytel{\reff@jnl{S\&T}}            
\def\solphys{\reff@jnl{Solar~Phys.}}    
\def\sovast{\reff@jnl{Soviet~Ast.}}     
\def\ssr{\reff@jnl{Space~Sci.Rev.}}     
\def\zap{\reff@jnl{ZAp}}                        
\def\nat{\reff@jnl{Nature}}             

\title[VSA First Results: Paper II]{First results from the Very Small
  Array -- II. Observations of the CMB}

\author[A C Taylor et al.] {Angela C. Taylor$^1$, Pedro Carreira$^2$,
  Kieran Cleary$^2$, Rod D. Davies$^2$, Richard J. Davis$^2$,
  \newauthor Clive Dickinson$^2$, Keith Grainge$^1$, Carlos M.
  Guti{\'e}rrez$^3$, Michael P. Hobson$^1$,\newauthor Michael E.
  Jones$^1$, R\"udiger Kneissl$^1$, Anthony Lasenby$^1$, J. P.
  Leahy$^2$, Klaus Maisinger$^1$, \newauthor Guy G. Pooley$^1$, Rafael
  Rebolo$^{3,4}$, Jose Alberto Rubi\~no-Martin$^3$, Ben
  Rusholme$^{1,\star}$, \newauthor Richard D. E. Saunders$^1$, Richard
  Savage$^1$, Paul F. Scott$^1$, An\v ze Slosar$^1$,\newauthor Pedro
  J. Sosa Molina$^3$, David Titterington$^1$, Elizabeth Waldram$^1$,
  Robert A.  Watson$^{2,\dagger}$, \newauthor and Althea
  Wilkinson$^2$.
  \\
  $^1$ Astrophysics Group, Cavendish Laboratory, University of Cambridge, UK.\\
  $^2$ Jodrell Bank Observatory, University of Manchester, UK.\\
  $^3$Instituto de Astrof{\'i}sica de Canarias, 38200 La
  Laguna, Tenerife, Spain.\\
  $^4$Consejo Superior de Investigaciones Cient{\'{\i}}ficas, Spain \\
  $^{\dagger}$Present address: Instituto de Astrof{\'{\i}}sica de
  Canarias.\\
  $^{\star}$Present address: Stanford University, Palo Alto, CA, USA}

\date{Accepted Received In original form}
\pagerange{\pageref{firstpage}--\pageref{lastpage}}
\pubyear{2002}

\begin{document}

\label{firstpage}
\maketitle

\begin{abstract}
  
  We have observed the cosmic microwave background temperature
  fluctuations in eight fields covering three separated areas of sky
  with the Very Small Array at 34~GHz. A total area of 101 square
  degrees has been imaged, with sensitivity on angular scales $3\fdg
  6$--$0\fdg 4$ (equivalent to angular multipoles $\ell$=150--900). We
  describe the field selection and observing strategy for these
  observations. In the full-resolution images (with synthesised beam
  of FWHM $\simeq 17$ arcmin) the thermal noise is typically
  $45\,\mu$K and the CMB signal typically $55\,\mu$k. The noise levels
  in each field agree well with the expected thermal noise level of
  the telescope, and there is no evidence of any residual systematic
  features. The same CMB features are detected in separate,
  overlapping observations. Discrete radio sources have been detected
  using a separate 15~GHz survey and their effects removed using
  pointed follow-up observations at 34~GHz.  We estimate that the
  residual confusion noise due to unsubtracted radio sources is less
  than $14$ mJy~beam$^{-1}$ ($15\,\mu$K in the full-resolution
  images), which added in quadrature to the thermal noise increases
  the noise level by 6 \%.  We estimate that the {\em rms}
  contribution to the images from diffuse Galactic emission is less
  than $6\, \mu$K. We also present images which are convolved to
  maximise the signal-to-noise of the CMB features and are co-added in
  overlapping areas, in which the signal-to-noise of some individual
  CMB features exceeds 8.

\end{abstract}


\section{INTRODUCTION}

\begin{figure*}
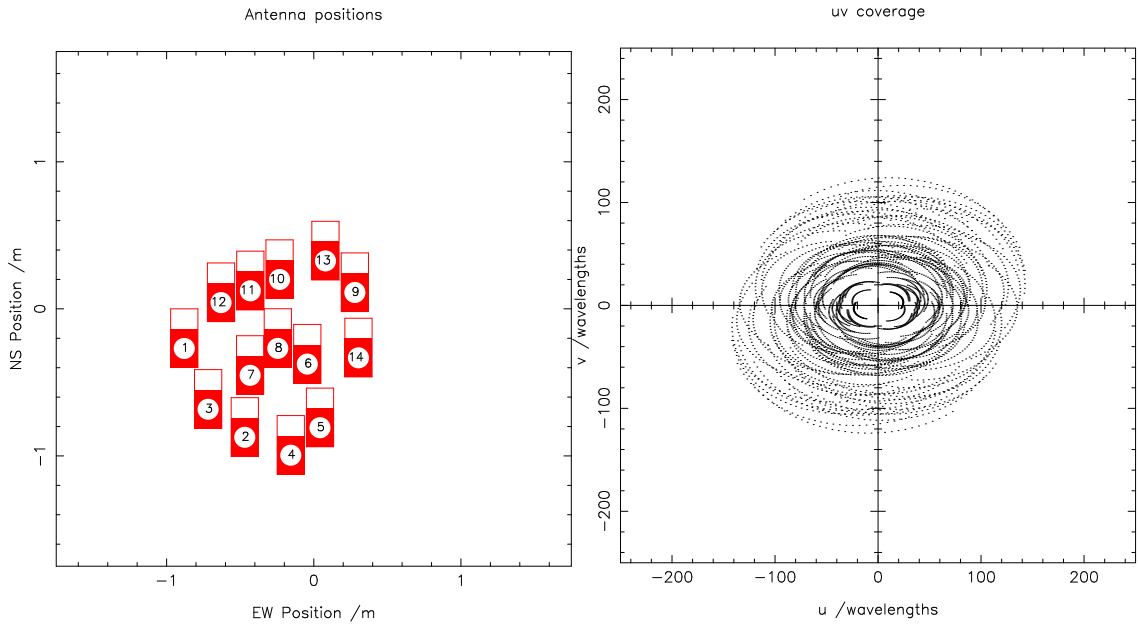

\begin{minipage}{150mm}
  \centering
  \epsfig{file=ant010101.ps,angle=-90,width=7.5cm}\epsfig{file=uv010101.ps,angle=-90,width=7.5cm}
\caption{Left: The array design used for the compact array.  Right:
  The resulting $uv$-coverage for a 6-hour observation.
  \label{fig:array_config}}
\end{minipage}
\end{figure*}

The measurement of primordial structure in the cosmic microwave
background (CMB) is of great importance in cosmology. These features,
imprinted at $z \sim 1000$, enable direct investigation of the
formation of structure in the universe as well as constraining the
values of basic cosmological quantities such as the amounts of various
forms of matter and vacuum energy in the universe. Primordial CMB
fluctuations, however, are extremely faint ($\Delta T < 100 \mu$K) and
have to be seen through various foregrounds and in the presence of the
systematic errors inevitably present in any experiment.

These challenges account for the gap of nearly three decades between
the discovery of the CMB \citep{Penzias65} and the detection of
primordial fluctuations by the COBE satellite \citep{Smoot92}. COBE
detected statistical anisotropy on angular scales of $90^\circ$ to
$10^\circ$ (corresponding to angular multipoles $\ell = 2$ to $20$);
this is greater than the horizon scale at the epoch of imprinting and
provided very strong evidence for a period of inflation in the early
universe. After COBE, observations have focused increasingly on the
angular scales at which features due to acoustic oscillations in the
primordial plasma were expected \citep{sakharov, sunyaev-70, peebles-70}.

Recently, results have been announced from a new generation of
instruments capable of detecting a range of acoustic modes. As well as
unequivocally detecting the first acoustic peak at $\ell \sim 220$,
MAXIMA~\citep{lee-01}, BOOMERanG~\citep{netterfield-01} and
DASI~\citep{halverson-02}, all detect power at smaller angular scales,
$300 < \ell < 800$, with strong evidence for a second peak at $\ell
\sim 550$, while CBI~\citep{cbi} (which is optimised for observation
at high $\ell$) detects the sharp decline in power at $\ell \sim 1000$
expected from photon diffusion~\citep{silk}.

Here we report measurements of the CMB anisotropies on scales of
$3\fdg 6$--$0\fdg 4$ (angular multipoles $\ell$=150--900) at
34~GHz using the Very Small Array (VSA). This paper is the second in a
series of four papers which report the results of the first season of
observations made using the VSA in its compact configuration. Here we
describe the observational strategy, data reduction and image-plane
analysis, whilst \citet{VSApaperI} (hereafter Paper I) provide a detailed
description of the experimental method and design.  Extraction of the
angular power spectrum is presented by \citet{VSApaperIII} (Paper III), and
estimates of the cosmological parameters using the VSA data are given by
\citet{VSApaperIV} (Paper IV). A detailed description of the telescope
will be given by Rusholme et al. (in prep.) (Paper 0).

\section{OBSERVATIONS}

\begin{table*}
\begin{minipage}{100mm}
 \caption{VSA field positions and total effective integration time
   remaining after flagging and filtering of the
   data.\label{tab:fieldpos}}
  \begin{tabular}{lccc}
    \hline
     & RA (J2000) & DEC (J2000)& Total Effective Integration Time (hrs)\\
       \hline
  VSA1  & 00 22 37.3    & 30 16 38 &  193\\
  VSA1A & 00 09 55.8    & 30 33 12 &  226\\
  VSA1B & 00 15 25.9    & 28 04 40 &  68\\
  VSA2  & 09 37 57.0    & 30 41 29 &  271\\
  VSA2-OFF & 09 47 15.2 & 30 41 05 &  226\\
  VSA3 & 15 39 39.3     & 44 50 21 &  187\\
  VSA3A & 15 30 35.1    & 42 37 49 &  206\\
  VSA3B & 15 46 20.8    & 42 22 15 &  119\\
  \hline
 \end{tabular}
\end{minipage}
\end{table*}

The VSA is a development of the Cosmic Anisotropy Telescope
(CAT)~\citep{cat_inst}, a three-element interferometer which operated
at 15~GHz from a sea-level site in Cambridge. The VSA has 14
horn-reflector antennas mounted on a tilt-table and operates over the
frequency range 26--36~GHz with an observing bandwidth of 1.5~GHz. The
telescope is sited at the Teide Observatory, Tenerife at an altitude
of 2400m.

Prior to our first season of CMB observations, a series of
commissioning observations was undertaken (see Paper I).  These
demonstrated that the telescope was working to specification and that
systematic effects such as the effects of the Sun and the Moon can all
be removed from the data to a very low level. Calibration of the data
was shown to be limited only by the uncertainty in the absolute flux
measurement of our primary calibrator, Jupiter, which is 3.5 \%; the
absolute flux calibration of VSA observations is based on the flux
scale of \cite{mason_casscal} and the brightness temperature of
Jupiter at 34~GHz is taken to be 154.5~K.

Detailed descriptions of the VSA, its data analysis procedures and
commissioning observations are given in Rusholme et al. (in prep.)
and Paper~I; in this paper we describe the key features of the first
season's observation of the CMB.

\subsection{Array configuration}

The 14 VSA antennas can be placed anywhere on the tilt-table, allowing
freedom to design the array configuration for specific observational
goals.  The `compact array' configuration used for the observations
reported in this paper was optimised to provide almost uniform
$uv$-coverage over the multipole range $\ell \sim150$ -- $900$, whilst
minimising the number of visibilities with low fringe rates. The sky
signal in such visibilities cannot be separated from the spurious
signal (see Paper I), and is predominantly contributed by the
shortest north-south baselines. Figure \ref{fig:array_config} shows
the VSA `compact array' configuration and $uv$-coverage.

\subsection{Observing frequency}

The VSA has the flexibility to observe anywhere in the 26--36~GHz
range, with 1.5~GHz instantaneous bandwidth. This tunability allows
the VSA to observe at more than one frequency. In principle, this
allows one to fit for a component of a known spectral index, i.e. it
would allow a separation of one or more Galactic components by virtue
of their spectral indices. We have in fact chosen to observe at just a
single frequency of 34~GHz. Our knowledge of the CMB foregrounds
suggests that the Galactic foregrounds at frequencies $\approx 30$~GHz
will contribute a few $\mu$K of signal in our fields (see section
\ref{galactic_foregrounds}). Choosing a frequency at the higher end of
the VSA observing band (34~GHz) reduces the free-free and synchrotron
foregrounds by a factor of $1.8-2.2$ compared to the lower end ($\sim
26$ GHz). Since the proposed spinning dust component is expected at
$\sim 15-20$ GHz \citep{draine-laz-diffuse}, its 34-GHz emission also
will be considerably lower than its peak value.

\subsection{Field selection}

For practical CMB observation, it is important to choose fields which are
relatively free from Galactic and extragalactic foregrounds.  All the VSA
fields are situated at Galactic latitudes greater than 20$^\circ$ and have low
Galactic synchrotron and free-free emission, as predicted by the 408-MHz
all-sky radio survey of \citet{408_mhz}.  The dust maps of \citet{schlegel-98}
were used to select fields with relatively low dust contamination. The actual
level of Galactic contamination in our fields is discussed in Section
\ref{galactic_foregrounds}.

To avoid bright clusters, we consulted the NORAS~\citep{noras},
BCS~\citep{rosat} and Abell~\citep{abell} catalogues.  All fields were
also chosen to be as free as possible of bright radio sources, since
these are the major contaminant of CMB observations at 34~GHz.  We
used two low-frequency surveys, NVSS~\citep{nvss} at 1.4~GHz and Green
Bank~\citep{gb6} at 4.85~GHz to select CMB fields in which there were
predicted to be no sources brighter than 500~mJy at 34~GHz within the
VSA primary beam (FWHM$=4\fdg 6$).  Predictions were made by
extrapolating the flux density of every source in the 4.8~GHz
catalogue to 34~GHz using its spectral index between 1.4 and 4.85~GHz.
A further practical consideration which affected the choice of CMB
fields was the need to observe all fields for a reasonable length of
time (more than 5 hours) from both Tenerife and Cambridge.  This
limits the declination range of our fields to between +26$^\circ$ and
+54$^\circ$. To maximize observing efficiency, we also selected fields
that are evenly spaced around the sky. In order to increase the
$\ell$-resolution of our measurement of the CMB power spectrum, we
selected regions of sky where we can mosaic several VSA pointings. In
each region of sky, mosaiced fields are separated by $2\fdg 75$.
Our final choice of fields used during the first year of observation
is shown in Figure~\ref{fig:fields} and in Table~\ref{tab:fieldpos}.

\subsection{Observing strategy}\label{strategy}

\begin{figure}
\centering
\psfig{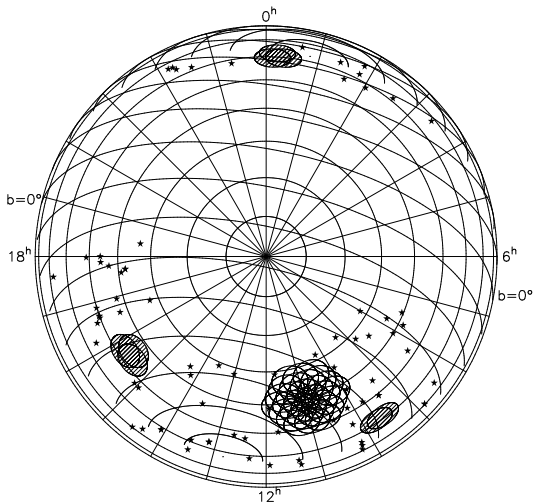}
\caption{Plot of the VSA fields observed during our first year of
  observations, projected on to the equatorial plane.  The hatched
  regions are fields which were observed using deep mosaicing, the
  results of which are the subject of this series of papers.  The
  remaining fields were observed as part of a shallow survey and are
  not discussed further here.  Radio sources predicted to be brighter
  than 500~mJy at 34~GHz are displayed as star symbols.  A cut-off at
  Galactic latitude $|\rm b|>$ 20$^{\circ}$ has been
  applied.\label{fig:fields}}
\end{figure}

During the first year of VSA observations, we undertook two distinct
observation programmes, each using the compact array.  First, we have
made deep mosaiced observations of eight fields in three evenly spaced
regions of sky (hatched regions in Figure \ref{fig:fields}).  Each
mosaiced field was observed for $\sim$~400 hours, reaching a thermal
noise of approximately 30~mJy.  Mosaicing in this way enables us to
increase the $\ell$-resolution of our measurements whilst also
reducing sample variance.  However, all the fields must be observed
with the Ryle Telescope in Cambridge for source subtraction (see
section 3.5), and the time taken for us to survey all fields with the
Ryle Telescope prior to any observation with the VSA has limited the
area of sky that we can cover with deep mosaicing in this first year.
Consequently, for $\ell\le$~300, our measurements of the CMB power
spectrum are limited by sample variance.

In order to achieve a good estimate of the CMB power spectrum in the
region $\ell\le$~300, we undertook a second observation programme; a
shallow survey in one large area of sky.  The shallow survey consists
of 2 days observation on each of 30 mosaiced fields (Figure
\ref{fig:fields} hexagonal region), and covers an area of
approximately 300 square degrees. For this shallow survey, our source
subtraction strategy no longer limits the area of sky we can observe
with the VSA in a given time.  Since for the shallow survey we are
only concerned with low-$\ell$ observations, where the contribution of
point sources to the CMB power spectrum will be negligible, prior
surveying with the Ryle Telescope is not necessary.  Instead, we
monitor only sources predicted to be greater than 100~mJy at 34~GHz on
the basis of low-frequency survey information.  Only the results from
the deep mosaiced fields are reported in this series of VSA results
papers; the results from the shallow survey will be presented in a
later paper.

\subsection{Observing schedule}

Quasi-independent tracking of the VSA antennas enables us to
effectively filter out the effects of bright contaminating sources
such as the Sun and the Moon (see Paper I).  We are able therefore to
observe during both the day and night, with useful CMB observations
possible at separations as close as 40~degrees from the Sun and
30~degrees from the Moon. Consequently, we implemented an observing
strategy that made use of all available time during the first year of
observation.

In each 24-hour period, we obtained 5--6 hours of integration on each
of 3 CMB fields; the total effective integration time (after flagging
and filtering of the data) for each of the VSA fields is given in
Table \ref{tab:fieldpos}. The remaining time in each 24-hour period
was divided between calibration sources. We observed at least one flux
calibrator (Jupiter, Cas A or Tau A), with observations of fainter
sources (e.g.  NGC7027, 3C273 and Cyg A) interspersed. Observation of
fainter sources allowed valuable checks on the diurnal stability of
the telescope.
 
\subsection{Data reduction}

The data from the first year of VSA observations were reduced as
described in detail in Paper I.  Each day's CMB observation is reduced
and analysed independently.  The resulting calibrated visibilities
from each day are stacked together, and the full dataset for each
field is analysed. We check the stacks by eye for signs of residual
contaminating signal, which is easily identified as regions of
coherent phase.  The data are stacked both as a function of hour angle
and per-antenna, and are flagged interactively.  This stage of
flagging typically removes 20$\%$ of the data; this is in addition to
the preliminary flagging operations described in Paper I, which
account for abnormal errors such as pointing, hardware failure, bad
weather and filtering of the spurious signal.
 
The fully-flagged and stacked data are held as visibility files,
containing the real and imaginary part for each observed $uv$-position
along with an associated rms noise level.  It is critical for the
power spectrum analysis that the estimate of the noise on each
visibility is accurate.  We calculate the noise on each visibility
point from the scatter in the visibility data on each baseline each
day.  This takes into account the number of flagged points in each
smoothed 64-second sample, in addition to other flagging operations
such as gain corrections implied by the system temperature monitoring.
As a further check, we recalculate our noise estimates when the data
are gridded in the $uv$-plane, prior to producing reduced datasets for
the power spectrum estimation.  Here we estimate the noise for each
gridded visibility using the spread of values within the stacked files
(i.e. using the complete 70--80 day data sets).  Both methods yield
the same noise estimates.

The final data stacks for each of the VSA fields typically contain
$\simeq$ 800,000 visibility points, each one averaged over 64 seconds.
The combined edits from all categories result in a rejection of $\simeq$
50$\%$ of the data in each field. 

\section{FOREGROUNDS}
\begin{table*}
\begin{minipage}{100mm}
\caption{Galactic foreground contamination for the 3 VSA mosaiced
fields at 1$^\circ$ resolution. The rms estimates are in $\mu$K, and
are calculated for 34~GHz.\label{tab:galactic_foreground}}
\begin{tabular}{llccc}
\hline
Component & Template / assumptions & VSA1 & VSA2 & VSA3\\ 
\hline
Synchrotron & Haslam 408~MHz     & 2.2 & 1.2 & 1.0\\
            & $\beta=3.0$  &     &     &    \\
Free-free   & WHAM H$\alpha$         & 0.8 & 0.8 & 0.6\\
            & 1R=4.5$~\mu$K         &     &     &    \\   
Spinning Dust & Schlegel et al. 100~$\mu$m    & 5.0 & 1.4 & 2.4\\
            & 1~MJy~sr$^{-1}$ = 10~$\mu$K &    &     &    \\ 
Total (estimated) & uncorrelated & 5.5 & 2.0 & 2.7\\
\hline
\end{tabular}
\end{minipage}
\end{table*}

\subsection{Diffuse Galactic foregrounds}

The diffuse Galactic foregrounds consist of three known components:
synchrotron, free-free and vibrational dust, together with a further
postulated component of spinning dust.

Synchrotron emission has a varying spectral index ($T \propto \nu^{-
  \beta}$) of $\beta_{synch} \approx 2.7-3.2$ \citep{davies-96}
depending on frequency and position on the sky. The steep spectral
index allows low-frequency radio surveys such as the 408-MHz all-sky
radio map \citep{408_mhz} and the 1420-MHz radio map \citep{reich-88}
to be used to estimate the synchrotron component at higher
frequencies.

Free-free emission is thermal bremsstrahlung radiation emitted by
ionized gas in the Galaxy. The spectral index for free-free is more
well-known with $\beta_{ff} \approx 2.15$; it shows very little
variation with electron temperature and density, although it does
steepen slightly at higher frequencies. The 656.3 nm H$\alpha$ line is
a good tracer of free-free emission (\cite{valls-gabaud-98} and
Dickinson et al. in prep). Dust absorption, however, can cause these
estimates to be systematically lower than the true values.

Vibrational dust is only a concern for CMB experiments observing at
frequencies above $\sim 90$ GHz \citep{dezotti-99}, and is expected to
contribute $<< 1~\mu$K at VSA observing frequencies.  A further
component has recently been postulated by \cite{draine-laz-diffuse},
based on observational evidence (\cite{kogut-high-lat, leitch-97,
  oliveira-97, oliveira-99, oliveira-02}, but see also
\cite{mukherjee-01}). They postulate a population of ultra-small
($\sim10^{-9}$m) spinning dust grains which emit via rotational
emission. The spectrum of spinning dust is likely to be highly peaked
at about $15-20$ GHz.

\subsection{Galactic foregrounds in the VSA fields}
\label{galactic_foregrounds}

A simple method of estimating the foreground contamination is to
calculate the rms variation in each of the foreground template maps
for each VSA field. This is then converted to temperature units at
34~GHz. This is simplistic and susceptible to uncertainty due to
the extrapolation to 34~GHz. This method, however, does give us an
estimate of the amplitude of the foreground signals and indicates to
what level the VSA observations are contaminated by diffuse
foregrounds. A more rigorous cross-correlation analysis will be
presented in a future paper (Dickinson et al., in prep.).

The rms method has been applied to $8^{\circ} \times 8^{\circ}$
patches of sky centred on each of the three VSA deep fields (VSA1,
VSA2 and VSA3). These are large enough to cover the mosaiced fields
within the $4\fdg 6$ FWHM of the primary beam. We use the
de-striped and source-subtracted version of the Haslam et al. (1982)
408 MHz map as used by \cite{davies-96} to trace the synchrotron
component, whilst the recently published WHAM H$\alpha$ data
\citep{reynolds-98} are used to trace the free-free component.  For
the dust-correlated emission, we use the 100 $\mu$m IRAS/DIRBE map
given by \cite{schlegel-98}.  The maps were smoothed to $1^{\circ}$
resolution since the WHAM data are at $\sim 1^{\circ}$ resolution.

The synchrotron emission is assumed to have a spectral index $\beta =
3.0$ between 408 MHz and 34 GHz. The conversion from H$\alpha$ units
(1 Rayleigh $(R) \equiv 10^{6}/4\pi $ photons s$^{-1}$ cm$^{-2}$
sr$^{-1}$) to free-free emission is calculated to be 4.5 $\mu$K/$R$
(Dickinson et al. in prep.). To estimate the $total$ dust-correlated
component (which may include emission from spinning dust and
vibrational dust) we have assumed a coupling coefficient of 10
$\mu$K/(MJy~sr$^{-1}$). This number is rather arbitrary since previous
estimates which range from $0-36 \mu$K/(MJy~sr$^{-1}$) at $\approx 30$
GHz, have been calculated for different angular scales and for
different regions of sky. The total value is the sum of the individual
components added in quadrature. The estimates of the components and
total contamination from Galactic foregrounds for the three fields are
given in Table~\ref{tab:galactic_foreground}.

The value of the synchrotron component is probably an over-estimate
since there are some residuals from discrete radio sources left in the
map; above $\sim$ 10~GHz, the free-free component is expected to
dominate the synchrotron component. The free-free estimate from
H${\alpha}$ is probably the most reliable although there may be some
absorption of the H$\alpha$ signal due to dust. Following Dickinson et
al. (in prep.), we can use the 100~$\mu$m emissivity to calculate what
this correction might be. For all the VSA fields, the dust emissivity
is $< 5$ MJy~sr$^{-1}$. If {\em all} the dust were lying in front of
the H$\alpha$ gas, then the correction is $23$ percent. A more likely
situation is a mixing of the gas and dust, thus the dust correction is
$\sim$ 10 percent. The table indicates the brightest Galactic
foreground may be the spinning dust, although very little is known
about this component. The estimates are likely to be higher than the
real value since we chose a rather high coupling coefficient based
primarily on the COBE-DMR dust-correlated value \citep{kogut-high-lat}
which is only strictly true at the 7$^{\circ}$ scale. This is the
total dust correlated component which will also include any other
correlated foreground, most notably, the free-free component.

In summary, using rms estimates for these VSA fields, it seems that
the diffuse Galactic foregrounds are very weak at 34 GHz compared to
the primordial CMB signal. For example, if $T_{\rm CMB}=60~\mu$K near
the first peak $\ell \sim 200$ and $T_{\rm GAL}=5~\mu$K, then the
contamination is at the $0.35$ per cent level. For $\ell \sim 800$, if
$T_{\rm CMB}=30~\mu$K, then the contamination is $1.4$ per cent.
However, the contamination is likely to be considerably smaller at
higher $\ell$-values, since all the Galactic foregrounds have power
spectra which decrease at smaller angular scales. The full
cross-correlation analysis (Dickinson et al., in prep.) will provide a
more accurate estimate of the contamination from Galactic foregrounds.

\subsection{Point sources}

Radio sources (mostly radio galaxies and quasars, but also some stars)
are expected to be a major contaminant of CMB observations at
centimetre wavelengths (see e.g., \citet{TGJ01}).  Their population is
not well known at frequencies higher than 10~GHz, and many are
expected to be variable and/or have spectra that rise with frequency,
i.e. $\alpha < 0$ (where flux density S $\propto \nu^{-\alpha}$).  On
the angular scales of interest to primordial CMB work, such sources
are also generally unresolved. To eliminate point source contamination
from VSA observations, all the radio sources in each CMB field with
flux density greater than our source-confusion limit, are observed and
subtracted directly from our visibility data.  We estimate the
contribution of unsubtracted sources below our confusion limit using a
preliminary estimate of the 34~GHz source count; this source count is
estimated from the sample of sources we detect at 34~GHz.

\begin{table}
  \caption{Point sources subtracted from VSA fields. The apparent flux
    is the flux of each source at 34~GHz as observed by the VSA
    (i.e. including the effect of the VSA primary beam).  This flux
    represents the actual flux subtracted from each map.\label{tab:sources}}
  \begin{tabular}{lllc}
\hline
       & RA (J2000) & DEC (J2000) & Apparent Flux (Jy)\\
\hline
  VSA1 & 00 10 27.7 & 28 54 58  &  0.028 \\
       & 00 15 06.1 & 32 16 13  &  0.155 \\
       & 00 19 39.7 & 26 02 53  &  0.043 \\
       & 00 23 39.6 & 29 28 29  &  0.091 \\
       & 00 24 34.3 & 29 11 31  &  0.088 \\
       & 00 28 17.0 & 29 14 29  &  0.054 \\
       & 00 34 43.4 & 27 54 26  &  0.037 \\
       & 00 36 59.7 & 29 58 59  &  0.029 \\
       & 00 41 15.9 & 32 11 08  &  0.011 \\
&&&\\  
VSA1A &  23 55 44.0 & 30 12 42    &  0.018\\
      &  00 10 27.7 & 28 54 58  &  0.058\\
      &  00 12 47.3 & 33 53 36  &  0.029\\
      &  00 13 44.0 & 34 41 42  &  0.008\\
      &  00 15 06.1 & 32 16 13  &  0.167\\
      &  00 15 37.9 & 33 32 41  &  0.017\\
      &  00 19 37.7 & 29 56 02  &  0.033\\
      &  00 23 39.6 & 29 28 29  &  0.019\\
      &  00 24 34.3 & 29 11 31  &  0.020\\
&&&\\
  VSA1B & 00 00 35.1 & 29 14 29 &   0.019\\
        & 00 00 55.9 & 25 16 24 &   0.014\\
        & 00 01 21.5 & 25 26 51 &   0.015\\
        & 00 03 16.3 & 24 59 42 &   0.008\\
        & 00 04 34.5 & 29 46 18 &   0.021\\
        & 00 06 48.8 & 24 22 38 &   0.027\\
        & 00 12 38.1 & 27 02 40 &   0.057\\
        & 00 15 06.1 & 32 16 13 &   0.033\\
        & 00 16 03.6 & 24 40 18 &   0.025\\
        & 00 18 12.4 & 29 21 25 &   0.101\\
        & 00 19 39.7 & 26 02 53 &   0.298\\
        & 00 19 52.4 & 26 47 31 &   0.052\\
        & 00 23 23.0 & 25 39 19 &   0.044\\
        & 00 34 43.4 & 27 54 26 &   0.015\\
&&&\\
  VSA2  & 09 23 48.0 & 31 07 56 &  0.047\\
        & 09 23 51.5 & 28 15 26 &  0.047\\
        & 09 25 43.6 & 31 27 11 &  0.062\\
        & 09 32 55.0 & 33 39 29 &  0.023\\
        & 09 39 01.6 & 29 08 29 &  0.055\\
        & 09 41 48.1 & 27 28 38 &  0.051\\
        & 09 52 06.1 & 28 28 31 & 0.011\\
        & 09 53 27.9 & 32 25 52  & 0.012\\
        & 09 58 20.9 & 32 24 02  & 0.014\\ 

\hline    
\end{tabular}
\end{table}
\begin{table}
  \contcaption{}
  \begin{tabular}{lllc}
\hline
       & RA (J2000) & DEC (J2000) & Apparent Flux (Jy)\\
\hline
   VSA2-OFF & 09 27 39.8 & 30 34 16  & 0.060\\
      & 09 31 51.8 & 27 50 52   &   0.013\\
      & 09 37 06.3 & 32 06 55   &   0.026\\
      & 09 41 48.1 & 27 28 38   &   0.035\\
      & 09 41 52.3 & 27 22 18   &   0.018\\
      & 09 42 36.1 & 33 44 37   &   0.017\\
      & 09 52 06.1 & 28 28 31   &   0.031\\
      & 09 53 27.9 & 32 25 52   &   0.053\\
      & 09 54 39.7 & 26 39 23   &   0.006\\
      & 09 55 37.9 & 33 35 03   &   0.018\\
      & 09 58 20.9 & 32 24 02   &   0.267\\
      & 09 58 58.9 & 29 48 04   &   0.026\\
      & 10 00 07.5 & 27 52 45   &   0.012\\
      & 10 01 10.1 & 29 11 38   &   0.043\\
      & 10 01 46.2 & 28 46 55   &   0.028\\
&&&\\
 VSA3 & 15 21 49.5 & 43 36 39  &   0.106\\
      & 15 39 42.4 & 43 37 41   &   0.089\\ 
      & 15 45 08.4 & 47 51 54  &   0.021\\
      & 15 56 36.1 & 42 57 07  &   0.018\\
      & 15 57 18.9 & 45 22 21  &   0.022\\
&&&\\
VSA3A & 15 06 52.9 & 42 39 22  &   0.056\\
      & 15 11 42.6 & 44 30 44  &   0.010\\
      & 15 16 31.4 & 43 49 49  &   0.023\\
      & 15 17 56.9 & 39 36 40  &   0.016\\
      & 15 19 26.9 & 42 54 08  &   0.066\\
      & 15 20 39.6 & 42 11 12  &   0.043\\
      & 15 21 49.5 & 43 36 39  &   0.238\\
      & 15 26 45.3 & 42 01 41  &   0.081\\
      & 15 36 13.7 & 38 33 27  &   0.014\\
      & 15 36 23.1 & 38 45 52  &   0.011\\
      & 15 38 55.7 & 42 25 27  &   0.065\\
      & 15 42 58.9 & 45 45 59  &   0.012\\
      & 15 45 21.3 & 41 30 27  &   0.026\\
      & 15 47 58.9 & 42 08 52  &   0.021\\
&&&\\
VSA3B & 15 21 49.5 & 43 36 39     &   0.017\\
      & 15 57 18.9 & 45 22 21     &   0.012\\
      & 16 08 22.0 & 40 12 15     &   0.010\\
&&&\\
\hline
  \end{tabular}
\end{table}

\subsection{Source confusion and the VSA}

The level to which sources must be subtracted from our CMB fields is
determined by comparing the flux sensitivity of the VSA to the rms
confusion noise remaining from unsubtracted sources.  The VSA,
operating in its compact configuration, has a flux sensitivity of
about 30~mJy in each deep mosaic pointing. In order that our
observations are dominated by receiver noise we must therefore ensure
that the rms flux density on a VSA map from unsubtracted sources, the
confusion noise, is less than the flux sensitivity of 30~mJy.
Following the standard approach (e.g. \cite{Scheuer}),
\[
 \sigma^{2}_{\rm conf}=\Omega\int_{0}^{S_{\rm sub}}n(S)S^{2}dS,
\]
where $S$ is the flux density, $n(S)$ is the differential source count
per steradian, $\Omega$ is the VSA synthesised beam solid angle and
$S_{sub}$ is the flux density limit down to which sources are
subtracted.  Normalising this in terms of the number of sources, $N$,
per VSA synthesised beam that are subtracted,
\[
N = \int_{S_{\rm sub}}^{\infty}\Omega n(S)dS,
\]
and parameterising $n(S)$ as $n(S)dS=KS^{-\beta}$, we obtain
\[
\sigma_{\rm conf}=[N(\beta-1)/(3-\beta)]^{1/2}S_{\rm sub}.
\]

For a given level of source subtraction, $S_{\rm sub}$, one can thus
estimate the residual confusion noise.  The level of source
subtraction can then be optimised to ensure that $\sigma_{\rm conf}$
is less than the flux sensitivity of the VSA.  To calculate this
confusion noise, however, the differential source count, $n(S)$, at
our observing frequency of 34~GHz is required.  At the outset of
observations with the VSA, the source count at 34~GHz was not known.
Instead we used an estimate extrapolated from the 15-GHz source count
obtained using source subtraction observations for the Cosmic
Anisotropy Telescope(CAT).  

Based on the small sample of sources observed during the CAT
source-subtraction program, the 15~GHz source counts were found to
have $\beta=-$1.8~\citep{osullivan}.  Using this value for $\beta$ at
15~GHz and assuming a conservative average spectral index of
$\alpha$=0 between 15 and 34~GHz, we find that to ensure that the
confusion noise is significantly lower than the flux sensitivity of
the VSA ($\simeq$~30 mJy) we must subtract all sources brighter than
80~mJy at 34~GHz.
\begin{table*}
\begin{minipage}{150mm}
\begin{center}
 \caption{This table gives the thermal noise level,in mJy beam$^{-1}$, for each
   VSA field, measured both from standard and autosubtracted maps (see
   text for explanation). The rms noise level in the centre of each
   map and the corresponding residual CMB signal observed in each
   field, in mJy, is also given.  These values are taken from CMB maps
   made at full resolution. The approximate flux-to-temperature
   conversion is $\rm 1 \, mJy \, beam^{-1} \equiv 1.1 \, \mu K$,
   giving mean rms CMB fluctuations in the maps of 40 --
   65~$\mu$K.\label{tab:noises} }
  \begin{tabular}{lcccc}
    \hline
       & Thermal Noise & Thermal Noise & RMS Noise & CMB Signal\\
        & (standard map)&(autosubtracted map)& in centre of map &\\
       \hline
  VSA1  & 37 & 38 & 52 & 37 \\
  VSA1A & 46 & 43 & 51 & 27 \\
  VSA1B & 60 & 63 & 81 & 55 \\
  VSA2  & 39 & 37 & 70 & 58 \\
  VSA2-OFF & 34   & 33 & 53  & 41  \\ 
  VSA3  & 39 & 39  & 56 & 40\\
  VSA3A & 36 & 39  & 63 & 52\\  
  VSA3B & 50 & 49  & 75 & 56\\ 
  \hline
 \end{tabular}
\end{center}
\end{minipage}
\end{table*}
\subsection{Source-subtraction strategy}

To ensure that all potentially contaminating sources are detected, one
would ideally survey each CMB field at the same frequency as the VSA,
but with increased resolution and sensitivity. This, however, would
require a telescope more expensive than the VSA.  Instead, we adopt a
two-stage strategy.
 
First, prior to observation with the VSA, we survey all the VSA fields
at 15~GHz using the Ryle Telescope (RT) in Cambridge. The RT, which
uses five 13-m diameter antennas and gives a resolution of $\simeq 30$
arcsecs, is used in a raster scanning-mode and reaches an rms noise
level of $\sigma \simeq 4$~mJy (see \citet{waldram-02}).  This allows
us to identify sources above 20~mJy at 15~GHz, and ensures that we
find all sources above our source-confusion limit of 80~mJy at 34~GHz,
even allowing for a spectral index as steep as $\alpha \simeq$~$-$2
between 15 and 34~GHz.

Having identified the contaminating sources in each field, we monitor
each source at 34~GHz using a separate single-baseline interferometer
working simultaneously with the VSA (see Paper I).  This
single-baseline interferometer consists of two 3.7-m dishes separated
by 9 metres on a north-south baseline. Each dish is situated in an
enclosure similar to that of the VSA, and is fitted with identical
horn-reflector feeds and receivers as used on the main array. All
sources identified in the 15~GHz survey are monitored.  The monitoring
is done simultaneously with and at the same frequency as the VSA
observation, ensuring that sources which are variable on time-scales
as short as a few days can be subtracted accurately.

On average, each source identified in the 15-GHz survey is observed
once every three days. The flux of each monitored source is then
averaged over the duration of each corresponding field observation
(typically 80 days) and its contribution is then subtracted directly
from the stacked visibility data.

\subsection{Point source observations}

In total, 346 sources with flux densities greater than 20~mJy were
identified in the 15~GHz survey of the VSA fields.  All of these
sources were observed with the source subtractor, and the list of
monitored point sources with flux densities greater than 60~mJy at
34~GHz is given in Table \ref{tab:sources}.  Although our
source-subtraction strategy requires that only sources brighter than
80~mJy must be subtracted, and our source list is only complete to
this level, since the data were available we subtracted point sources
down to a flux limit of 60~mJy.  Typically, between 10 and 15 sources
with flux density greater than 60~mJy were subtracted from each CMB
field.

To calculate the residual point-source contamination remaining after
source-subtraction, we used a preliminary source count at 34~GHz.  The source
count is based on the 78 sources with $S_{34} > 60$~mJy detected by the source
subtractor. The differential count is given approximately by
\[
\frac{{\rm d}N}{{\rm d}S} = 54 \left( \frac{S}{\rm Jy} \right)^{-2.15}
\, {\rm Jy}^{-1}{\rm sr}^{-1}.
\]

Note that this count is not entirely accurate due to the uncertain
completeness at lower flux levels, but is sufficient to estimate our residual
confusion noise. A more complete analysis of the 34-GHz count will be given in
a subsequent paper. When the count is integrated up to our nominal subtraction
limit of 80~mJy it gives a residual confusion noise of $\sigma_{\rm conf} =
14$~mJy, which when added in quadrature to our typical thermal noise of 40~mJy
gives only a 6~\% increase in noise level. The effect of the residual sources
on the CMB power spectrum is discussed in Paper III.

\section{RESULTS}

\begin{figure*}
\begin{minipage}{150mm}
\epsfig{file=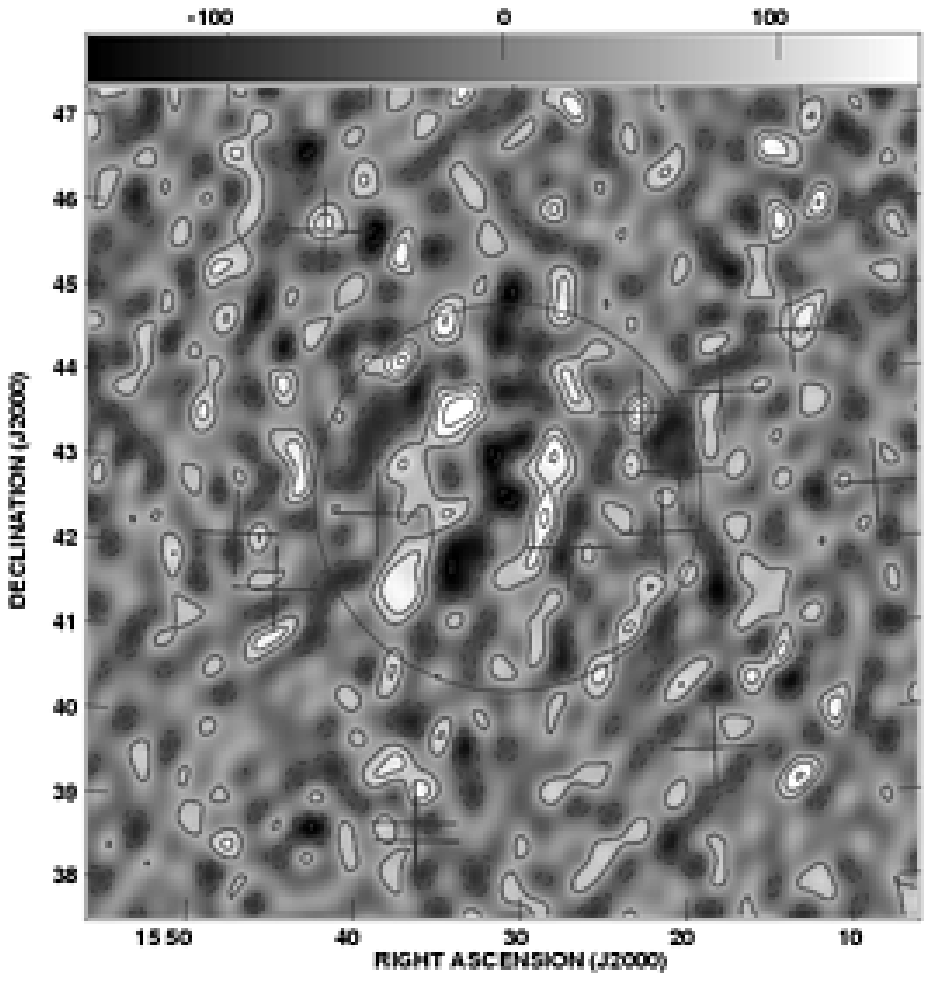,width=7.5cm}
\epsfig{file=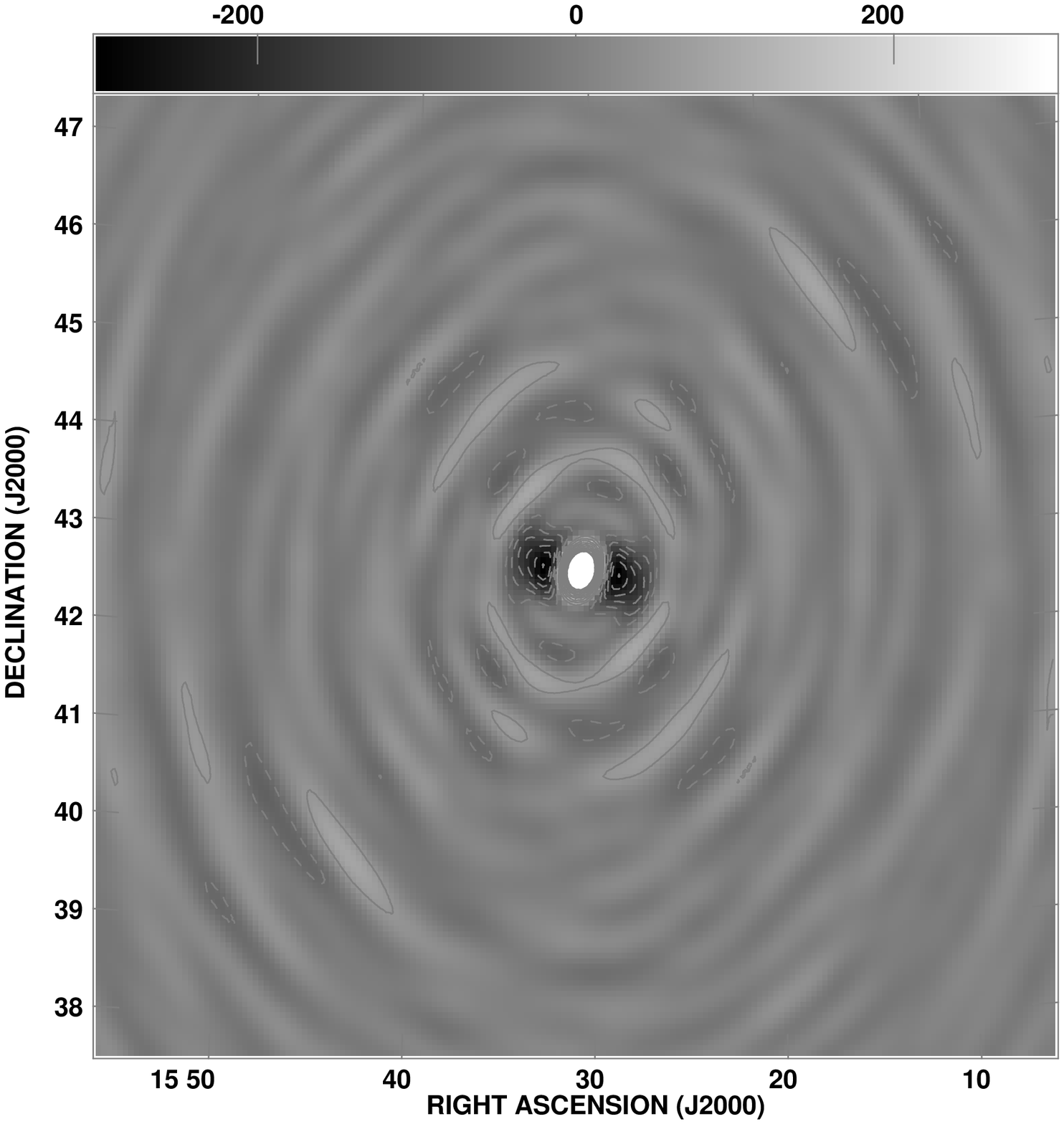,width=7.5cm}
\epsfig{file=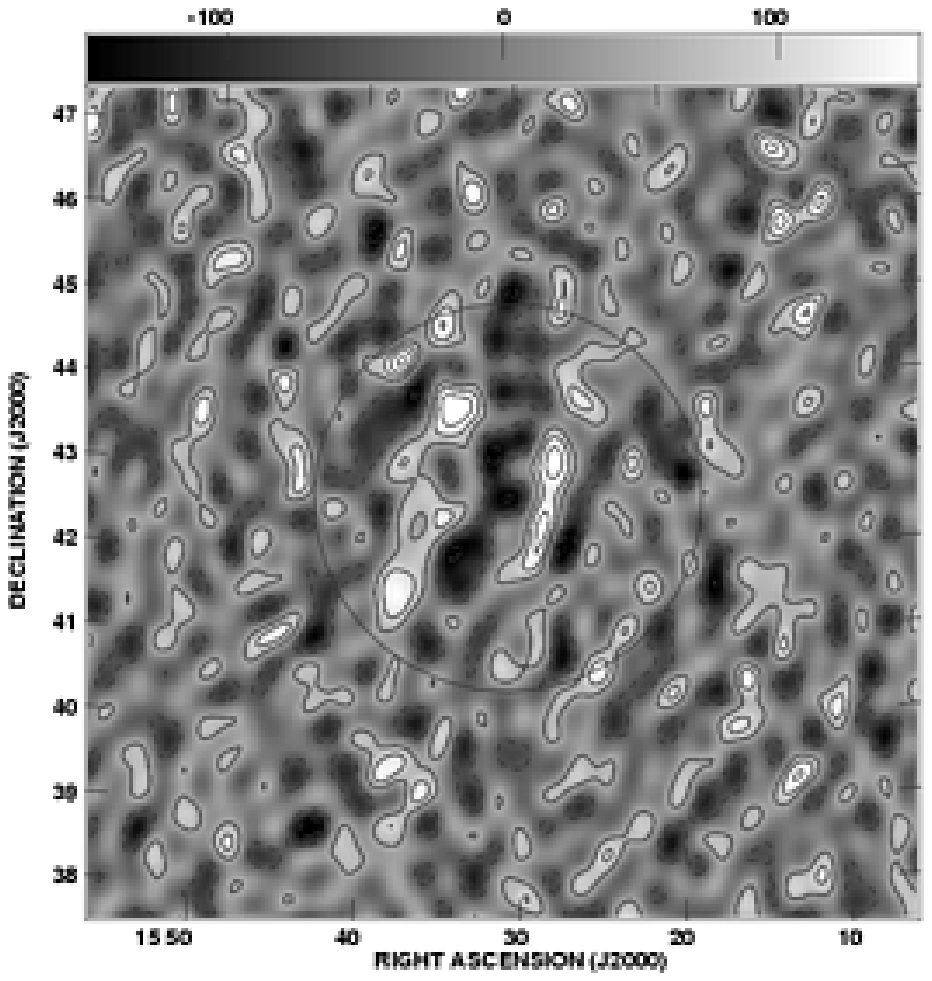,width=7.5cm}
\epsfig{file=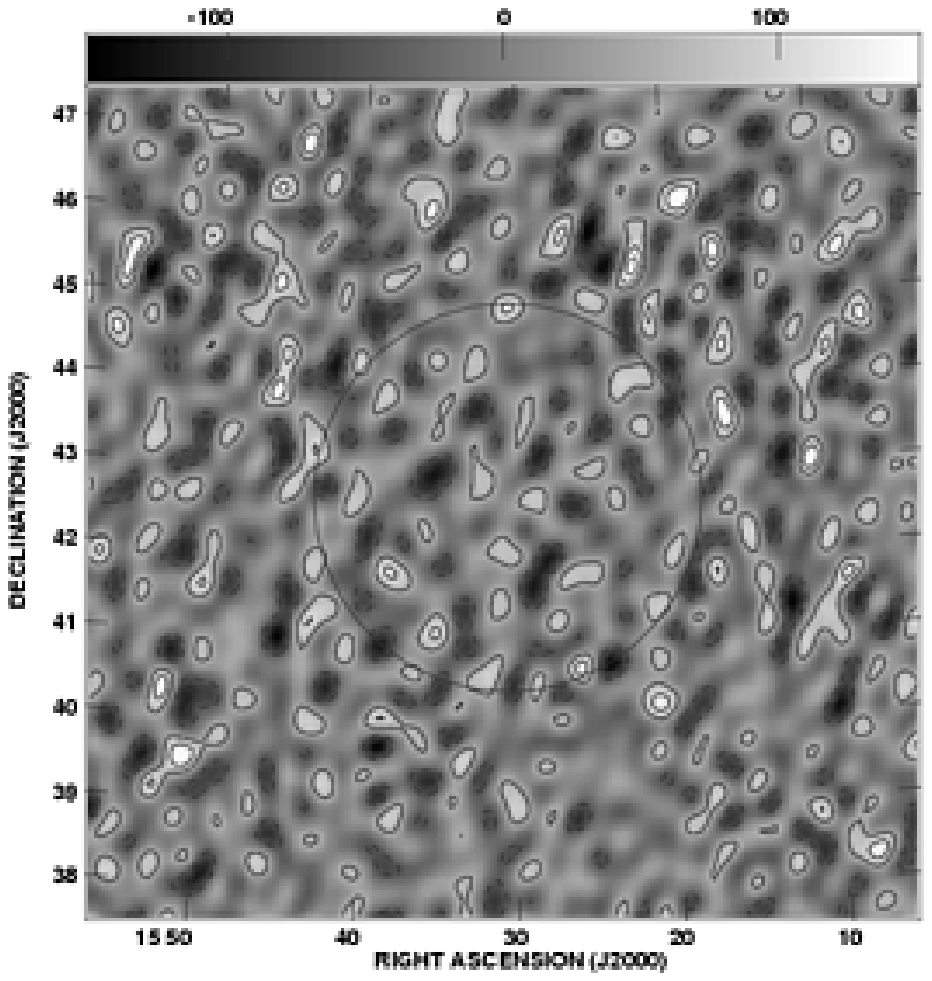,width=7.5cm}

\caption{Maps of a sample VSA field (VSA3A) at full resolution. Top left: the
  map before source subtraction. Positions of sources to be subtracted
  are marked with crosses, and the FWHM of the primary beam is
  indicated by a circle. CMB fluctuations (and some sources) are
  clearly visible within the primary beam. Top right: the synthesised
  beam of all the images. Bottom left: the map after source
  subtraction. Bottom right: the autosubtracted map (see text),
  showing the uniform noise level across the map. In all maps the
  greyscale runs from -150 to 150 mJy~beam$^{-1}$ and the contour
  interval is 50 mJy~beam$^{-1}$; in the beam the greyscale is -0.3 to
  0.3 and the contour interval is 5\% of the
  peak.\label{fig:sample_maps}}
\end{minipage}
\end{figure*}

Although the CMB power spectrum is calculated directly from the complex
visibility data obtained from our observations (see Paper III), image-plane
analysis provides valuable consistency checks on the data.  Here we present
images of the VSA fields and discuss the checks that we have performed on the
data.

Figure \ref{fig:sample_maps} shows images of a typical VSA individual
field (VSA3A) at full resolution.  The top two panels show the map
before source-subtraction and the synthesised beam, while the lower
left panel shows the map after source subtraction. Inside the primary
beam envelope (FWHM = 4\fdg 6, shown by a circle), the maps are
dominated by astronomical signal. Outside this region, the maps are
dominated by instrumental (thermal) noise.

Comparison of the rms power measured in the outer region of each map with the
calculated instrumental noise provides information about the level of residual
contaminating signals. Unwanted signals (such as crosstalk or distant bright
sources) are not constrained to lie within the primary beam envelope in the
image plane.  Instead their effect is to increase the total rms power level
across the whole of a map. A simple way to measure the instrumental noise
directly from a map is to measure the rms level far from the primary beam.
The measured noise levels, given in Table \ref{tab:noises}, all agree
extremely well with the calculated values.  The slightly higher noise levels
obtained in VSA1B and VSA3B reflect the smaller amount of data collected on
these fields.

A more rigorous approach to estimating the thermal noise on each map, and also
the amount of any residual contaminating signal, is to `autosubtract' our
visibility data.  Here we take the time-ordered visibility data for each field
and reverse the sign of alternate visibilities.  On the time scale between
adjacent visibility points (64 seconds), both astronomical and spurious
signals are effectively coherent, and are thus cancelled out.  In contrast,
the noise on adjacent visibilities is completely uncorrelated, and the rms
noise on the map is unaffected. This technique therefore provides a robust
estimate of our thermal noise. In addition, we can compare the rms power in
the autosubtracted maps with that far outside the primary beam envelope of our
standard field maps.  Any discrepancy between the two would be indicative of
residual spurious signals.

Autosubtraction was applied to all eight VSA fields, and maps of each field
were made.  The bottom right panel in Figure \ref{fig:sample_maps} shows the
autosubtracted map for VSA3A.  Clearly the astronomical signal has been
subtracted, leaving a constant noise level. The rms noise on each
autosubtracted map was measured and the values are given in Table
\ref{tab:noises}.  There is excellent agreement between the noise levels
measured far from the map centres and those measured by autosubtraction.

The approximate CMB signal level in each map can be estimated by subtracting
the map rms in the central area ($4 \times 4$ degrees) in quadrature from the
rms noise level estimated from the autosubtracted maps (or from the far-out
portions of the sky maps). These figures are also given in Table
\ref{tab:noises}, and represent the average temperature fluctuations in the
CMB averaged over the $\ell$-range of the observations. For the synthesised
beam size of the full resolution maps ($\simeq 17$~arcmin), the approximate
flux-to-temperature conversion is $\rm 1 \, mJy \, beam^{-1} \equiv 1.1 \, \mu
K$, giving mean rms CMB fluctuations in the maps of 40 -- 65~$\mu$K.

All the maps are robust to minor variations in flagging and filtering.
The data have also been split by observing epoch, by day/night and by
alternate visibilities; the resulting maps are fully consistent with
each other.  Analogues of theses consistency checks are also applied
to the power spectra (see Paper III).
\begin{figure}
\centering
\psfig{figure=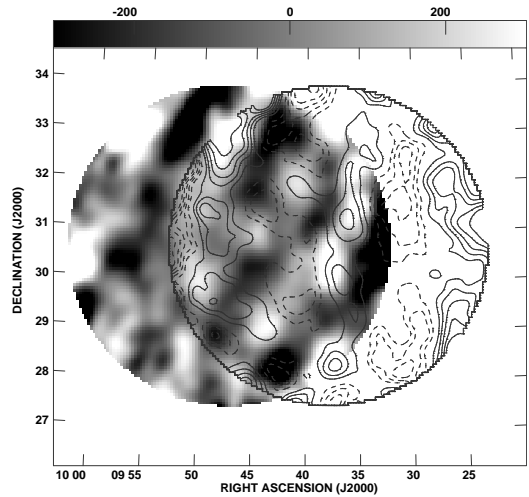,angle=270,width=7cm}
\caption{Comparison of the overlapping region in the VSA2 mosaic. A contour
map of VSA2-OFF is overlayed on a greyscale image of VSA2. Structures common
in both position and flux level are visible. Both maps were made with a taper
corresponding to the CMB power spectrum to maximise the signal-to-noise,
have been corrected for the primary beam response of the VSA, and are
truncated at the 25\% point of the primary beam. The grey scale runs from
$-300$ to $300$~mJy~beam$^{-1}$ and the contour interval is 150
mJy~beam$^{-1}$.\label{fig:overlap_map}}
\end{figure}

In addition to analysing each field independently, it is instructive to
compare images of fields within each mosaiced region.  As an example, we
present an image of the VSA2/VSA2-OFF mosaic (Figure
\ref{fig:overlap_map}). Here a contoured map of VSA2-OFF is overlaid on a
grey-scale image of the VSA2 field.  Common structures are clearly seen,
agreeing in both position and in flux density.

\begin{figure*}
\begin{minipage}{150mm}
\centering
\epsfig{figure=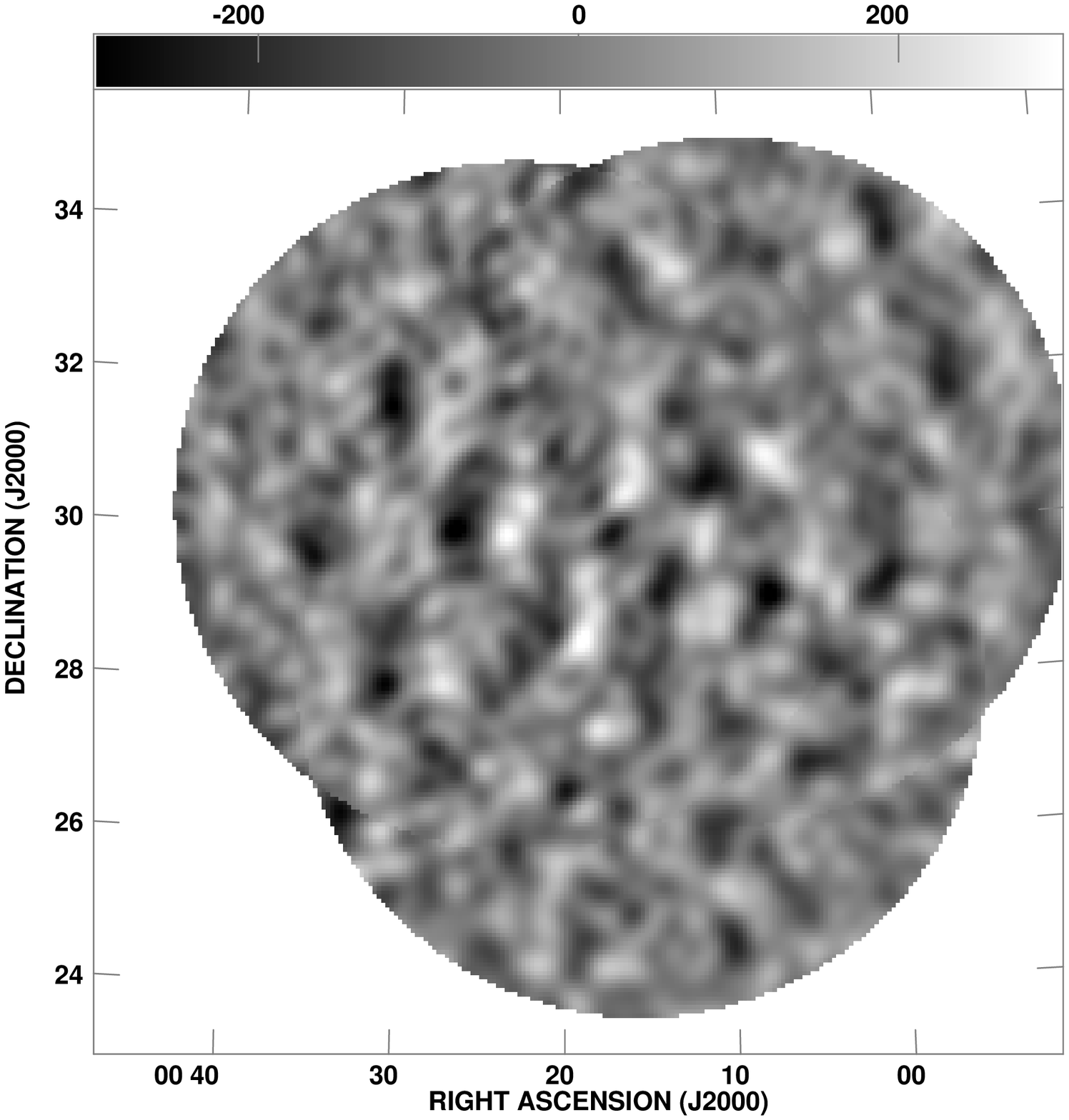,height=7.5cm}
\epsfig{figure=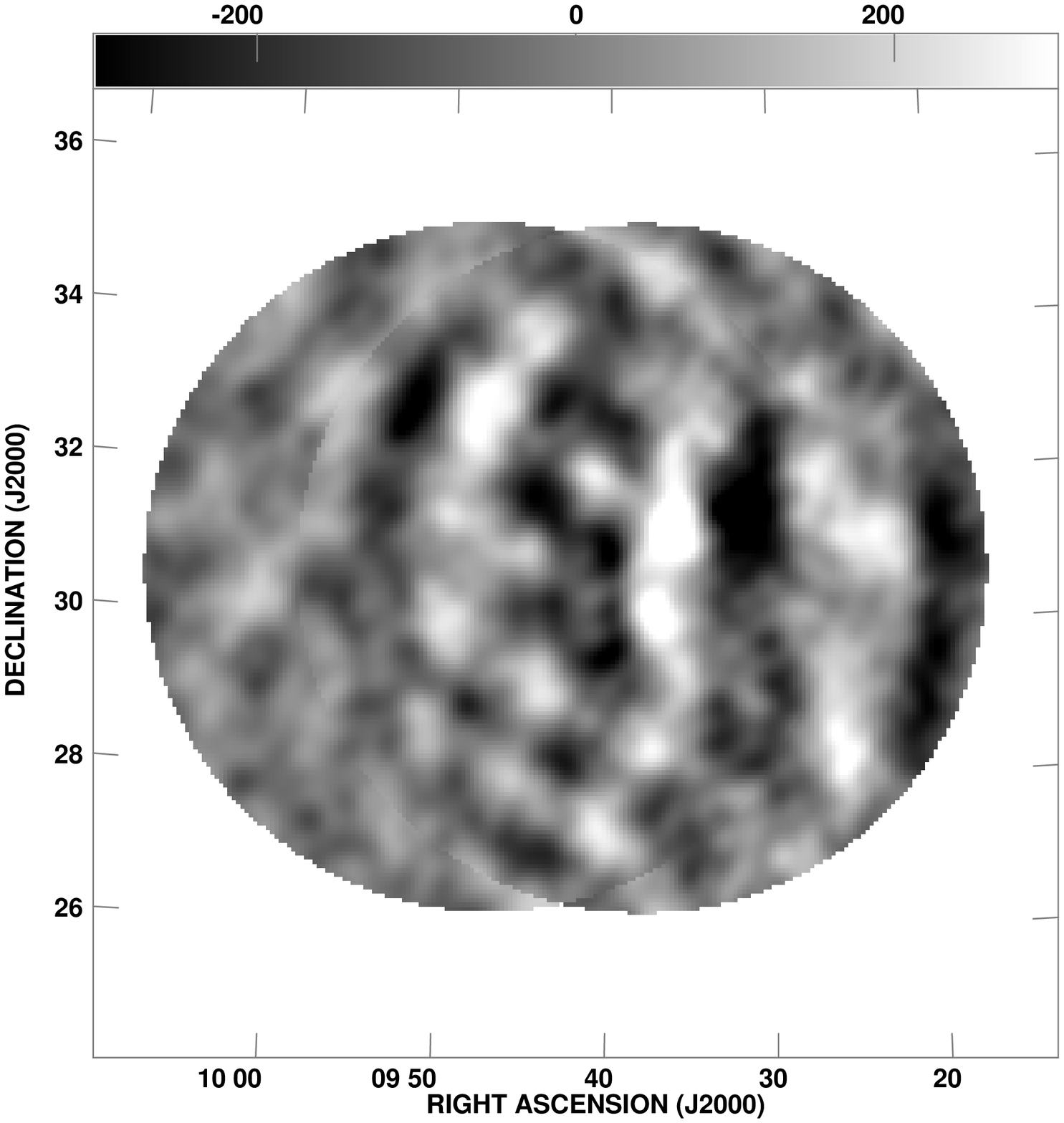,height=7.5cm}\\
\epsfig{figure=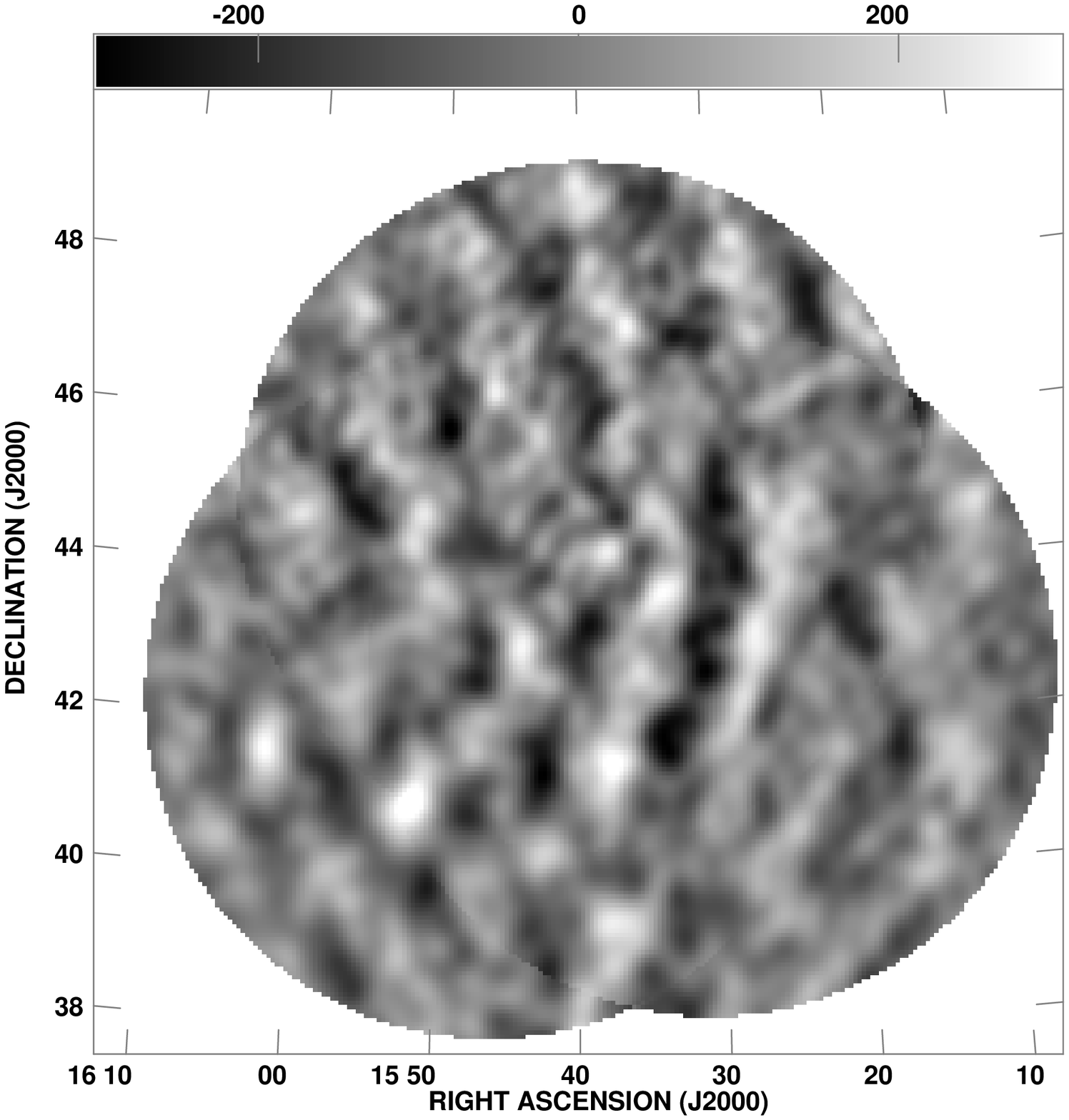,height=7.5cm} 
\epsfig{figure=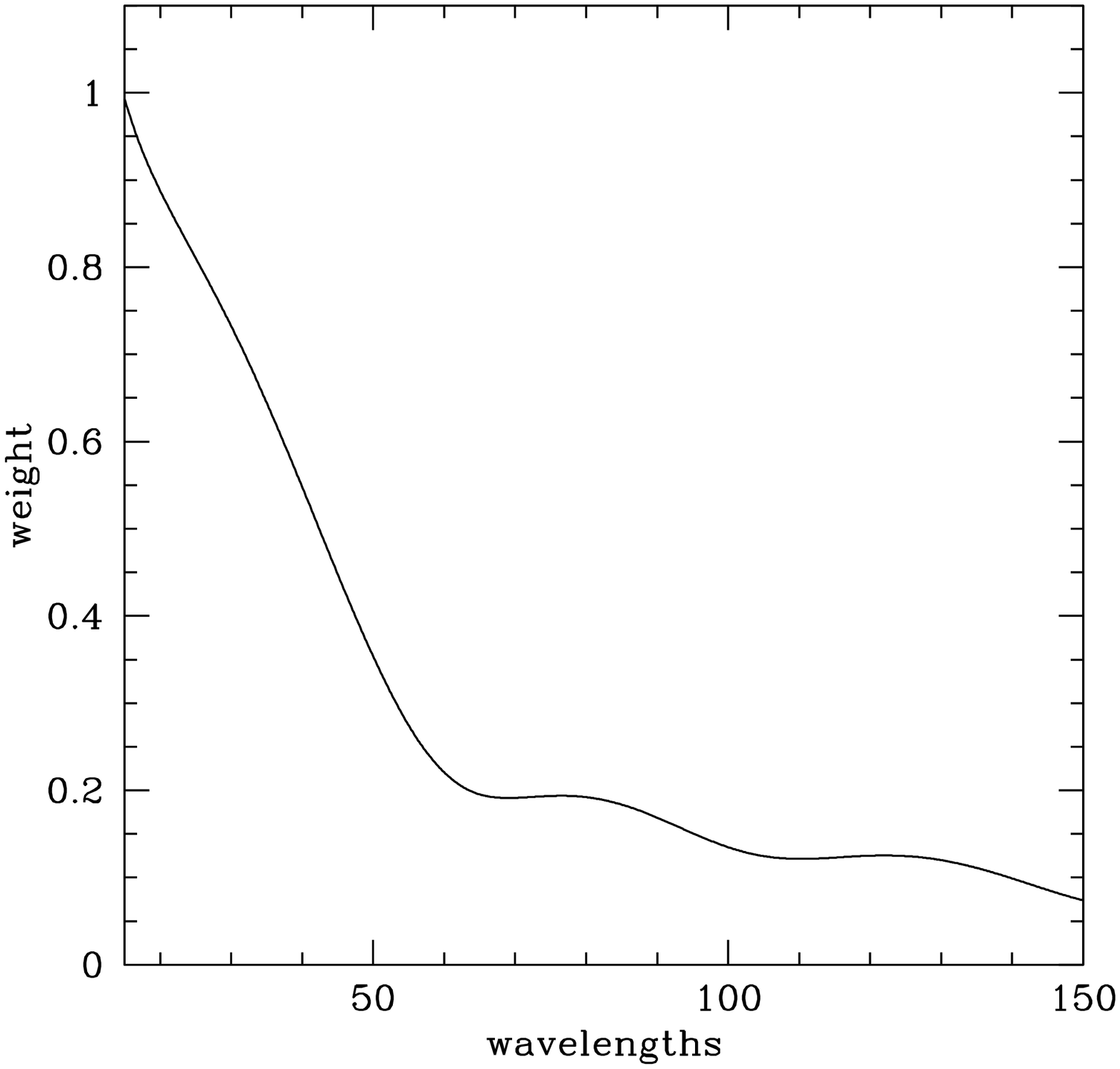,height=7.5cm}
\caption{Images of the three VSA regions, with radio sources
  subtracted. The individual maps were made with a radial taper
  applied to the visibilities in order to enhance the signal-to-noise
  of the CMB features, then combined using the {\sc Aips} tasks {\sc
    ltess} and {\sc stess}. These form a linear combination of the
  component maps, weighted by the individual primary beam responses.
  The point-spread function (or {\em synthesised beam}) have not been
  deconvolved from these images, which thus have correlations across
  them. Differing aperture plane coverages due to different flagging
  of the data have resulted in different synthesised beams in each
  image. Top left: VSA1,1A and 1B. Top right: VSA2 and VSA2-OFF.
  Bottom left: VSA3,3A and 3B. In each case the greyscale runs from
  -300 to 300 mJy ~beam$^{-1}$. Bottom right: The weighting function
  used to taper the maps in order to improve the signal-to-noise
  ratio, based on the power spectrum of the data; the weighting
  function is proportional to $C_{\ell}^{1/2}$. \label{fig:mosaic_maps} }
\end{minipage}
\end{figure*}

Finally, in Figure \ref{fig:mosaic_maps}, we present images of the
fully source-subtracted VSA fields. To enhance signal-to-noise, a
$uv$-taper function has been applied to the data.  The taper has the
form of the CMB power spectrum estimated from the complete VSA data
set (see Paper III). The individual overlapping maps are combined
linearly using standard {\sc {Aips}} tasks. The individual maps are
transformed to the same grid, then corrected for the individual
primary beam responses (which multiplies up the outer parts of the
map); they are then combined pixel by pixel, weighted by their primary
beams, and the resulting image multiplied by the total weight map,
resulting in an image in units of signal-to-noise. This is then
re-scaled back to the map units of Jy beam$^{-1}$. (This final
re-scaling is only approximate as the sensitivities vary slightly
between the component maps; however, no quantitative analysis is done
using these images). The peak value of the signal-to-noise varies
between 6 and 8.5 in the three combined images. The maps have not been
CLEANed, so the CMB fluctuations are convolved with the synthesised
beam, which is somewhat different for each field due to differences in
flagging.

\section{CONCLUSIONS}

We have observed eight overlapping fields in three separate regions of
the sky with the VSA compact array, a total of 101 square degrees. The
CMB anisotropies are clearly detected in all fields, and individual
features in different overlapped pointings agree well with each other.
We have assessed the Galactic foreground contamination based on
low-frequency radio, dust, and ${\rm H}_{\alpha}$ data, and found it
to be negligible. To eliminate confusion by foreground radiosources,
we have surveyed all the fields at 15~GHz with the Ryle Telescope
(RT).  The flux of all sources found in the RT survey were monitored
simultaneously with and at the same frequency as the VSA observations
using a separate single-baseline interferometer. These sources were
subtracted from the data. The count of sources in the VSA fields is
effectively complete to 80~mJy at 34~GHz, although we have detected
sources as faint as 60~mJy. After removal of these sources from the
VSA main array data, the residual source confusion is negligible. We
have checked the data for evidence of components other than the CMB
and thermal noise, and have found none.

\section*{ACKNOWLEDGEMENTS}

We thank the staff of the Mullard Radio Astronomy Observatory, Jodrell
Bank Observatory and the Teide Observatory for invaluable assistance
in the commissioning and operation of the VSA. The VSA is supported by
PPARC and the IAC. Partial financial support was provided by Spanish
Ministry of Science and Technology project AYA2001-1657.  A. Taylor,
R. Savage, B. Rusholme, C. Dickinson acknowledge support by PPARC
studentships. K. Cleary and J. A. Rubi\~no-Martin acknowledge Marie
Curie Fellowships of the European Community programme EARASTARGAL,
``The Evolution of Stars and Galaxies'', under contract
HPMT-CT-2000-00132. K. Maisinger acknowledges support from an EU Marie
Curie Fellowship. A. Slosar acknowledges the support of St. Johns
College, Cambridge. We thank Professor Jasper Wall for assistance and
advice throughout the project.

\bibliography{paper2}
\bibliographystyle{mn2e}

\bsp 

\label{lastpage}

\end{document}